\documentclass[twoside]{article}
\usepackage{amssymb}

\usepackage{fleqn,espcrc2}
\usepackage{graphicx}
\usepackage{amsmath}

\begin{document}

\title{D-XY Critical Behavior in Cuprate Superconductors}
\author{T. Schneider and J. M. Singer \\
\bigskip {Physik-Institut, Universit\"{a}t Z\"{u}rich,}\\
Winterthurerstr. 190, CH-8057 Z\"{u}rich, Switzerland}

\begin{abstract}
We outline the universal and finite temperature critical properties of the $%
3D$-$XY$ model, extended to anisotropic extreme type-II superconductors, as
well as the universal quantum critical properties in $2D$. On this basis we
review: (i) the mounting evidence for $3D$-$XY$ behavior in optimally doped
cuprate superconductors and the $3D$ to $2D$ crossover in the underdoped
regime; (ii) the finite size limitations imposed by inhomogeneities; (iii)
the experimental evidence for a $2D$-$XY$ quantum critical point in the
underdoped limit, where the superconductor to insulator transition occurs;
(iv) the emerging implications and constraints for microscopic models.
\end{abstract}

\maketitle

The starting point of the phenomenological theory of superconductivity is
the Ginzburg-Landau Hamiltonian {\small
\begin{eqnarray}
\mathcal{H}&=&\int d^{D}\mathbf{R}\bigg( \sum\limits_{j=1}^{D}{\frac{\hbar
^{2}}{2M_{j}}} \left|\left(i\nabla _{j}+{\frac{2\pi }{\Phi _{0}}}A_{j}(%
\mathbf{R})\right) \Psi\right|^{2}  \notag \\
&-&r|\Psi |^{2}+{\frac{u}{2}}|\Psi |^{4}+{\frac{|\mathrm{rot} \mathbf{A}|^{2}%
}{8\pi }}\bigg), \ j=(x,y,z).  \label{HP1}
\end{eqnarray}
} $D$ is the dimensionality of the system, the complex scalar $\Psi \left(
\mathbf{R}\right) $ is the order parameter, $M$ the effective mass of the
pair and $\mathbf{A}$ the vector potential. The pair carries a non-zero
charge in addition to its mass. The charge ($\Phi _{0}=hc/2e$) couples the
order parameter to the electromagnetic field via the first term in $\mathcal{%
H}$.

If $\Psi $ and $\mathbf{A}$ are treated as classical fields, the relative
probability $\mathcal{P}$ of finding a given configuration $[\Psi,\Psi^{\ast
},\mathbf{A}]$ is then {\small
\begin{eqnarray*}
\mathcal{P}[\Psi,\Psi^{\ast },\mathbf{A}] &=&\exp(-\beta\mathcal{H}%
[\Psi,\Psi^{\ast },\mathbf{A}]), \\
\beta&=&1/(k_{B}T).
\end{eqnarray*}
} The free energy $F$ follows from {\small
\begin{equation}
\exp\left(-{\frac{F}{k_{B}T}}\right)=Z=\int D[\mathbf{A}]D[\Psi]D[\Psi^{%
\ast}] \mathcal{P}[\Psi,\Psi^{\ast },\mathbf{A}],  \label{HP2}
\end{equation}
} where the partition function on the right hand side corresponds to an
integral over all possible realizations of the vector potential $\mathbf{A}$%
, the order parameter $\Psi$ and its complex conjugate $\Psi ^{\ast }$.
Setting $e=0$, the free energy reduces to that for a normal to neutral
superfluid transition, which is one of the best understood continuous phase
transitions with unparalleled agreement between theory, simulations and
experiment \cite{Ahlers:1984}. In extreme type-II superconductors, however,
the coupling to vector potential fluctuations appears to be weak \cite
{FFH:1991}, but nonetheless these fluctuations drive the system -- very
close to criticality -- to a charged critical point \cite
{Herbut:1996,Nguyen:1999}. In any case, inhomogeneities prevent cuprate
superconductors from entering this regime, due to the associated finite size
effect. For these reasons, the neglect of vectorpotential fluctuations
appears to be a reasonable starting point. In this case the vectorpotential
in Hamiltonian Eq. (\ref{HP1}) can be replaced by its most probable value.
The critical properties at finite temperature are then those of the $3D$-$XY$
model, reminiscent to the lamda transition in superfluid helium, but
extended to take the effective mass anisotropy into account \cite
{SchneiderAriosa:1992,ToniHugo:1993}.

The universal properties of the $3D$-$XY$ universality class are
characterized by a set of critical exponents, describing the asymptotic
behavior of the correlation length $\xi _{i}^{\pm }$, magnetic penetration
depth $\lambda _{i}$, specific heat $A^{\pm }$, etc., in terms of {\small
\begin{equation}
\xi _{i}^{\pm }=\xi _{i,0}^{\pm }|t|^{-\nu },\ \lambda _{i}=\lambda
_{i,0}|t|^{-\nu /2},\ C=\frac{A^{\pm }}{\alpha }|t|^{-\alpha },  \label{HP3}
\end{equation}
} where $3\nu =2-\alpha $. As usual, in the above expression $\pm $ refer to
$t=T/T_{c}-1>0$ and $t<0$, respectively. The critical amplitudes $\xi
_{i,0}^{\pm }$, $\lambda _{i,0}^{2}$, $A^{\pm }$, etc., are nonuniversal,
but there are universal critical amplitude relations, including \cite
{ToniHugo:1993} {\small
\begin{eqnarray}
(k_{B}T_{c})^{3} &=&\left( {\frac{\Phi _{0}^{2}}{16\pi ^{3}}}\right) ^{3}{%
\frac{\xi _{x,0}^{-}\xi _{y,0}^{-}\xi _{z,0}^{-}}{\lambda _{x,0}^{2}\lambda
_{y,0}^{2}\lambda _{z,0}^{2}}}  \notag \\
&=&\left( {\frac{\Phi _{0}^{2}}{16\pi ^{3}}}\right) ^{3}{\frac{\left(
\mathcal{R}^{-}\right) ^{3}}{A^{-}\lambda _{x,0}^{2}\lambda
_{y,0}^{2}\lambda _{z,0}^{2}}},  \notag \\
(\mathcal{R}^{\pm })^{3} &=&{A}^{\pm }\xi _{x,0}^{\pm }\xi _{y,0}^{\pm }\xi
_{z,0}^{\pm }.  \label{HP4}
\end{eqnarray}
} The singular part of the free energy density adopts in an applied magnetic
field $H$ the scaling form \cite{Schneider:1998_1} {\small
\begin{equation}
f_{s}={\frac{k_{B}TQ_{3}^{\pm }}{\xi _{x}^{\pm }\xi _{y}^{\pm }\xi _{z}^{\pm
}}}G_{3}^{\pm }({\mathcal{Z}}),\quad G_{3}^{\pm }(0)=1,  \label{HP6}
\end{equation}
} where {\small
\begin{eqnarray}
\mathbf{H} &=&H(\cos (\phi )\sin (\delta ),\sin (\phi )\sin (\delta ),\cos
(\delta )),  \notag \\
\mathcal{Z} &=&{{\frac{1}{\Phi _{0}}}\sqrt{H_{x}^{2}\xi _{y}^{2}\xi
_{z}^{2}+H_{y}^{2}\xi _{x}^{2}\xi _{z}^{2}+H_{z}^{2}\xi _{x}^{2}\xi _{y}^{2}}%
}.  \label{HP7}
\end{eqnarray}
} $\mathcal{R}^{\pm }$ and $Q_{3}^{\pm }$ are universal numbers, and $%
G_{3}^{\pm }(\mathcal{Z})$ is an universal scaling function of its argument.

\begin{figure}
\centerline{\includegraphics[height=5.2cm]{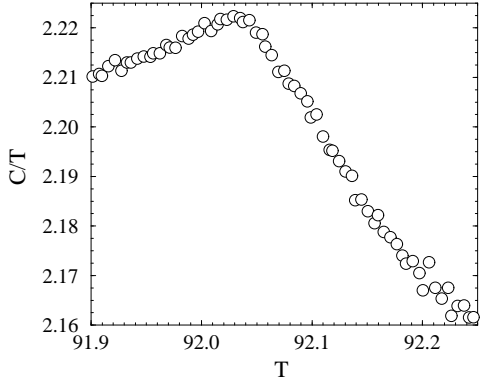}} \vskip -1.3cm
\caption{Specific heat coefficient $C/T$ $[mJ/(gK^{2})]$ versus $T$ $[K]$ of
$\mathrm{YBa_{2}Cu_{3}O_{7-\protect\delta}}$ (sample YBCO3, \protect\cite
{Charalambous:1999}). The two arrows mark $T_{c}\approx 92.12K$ and $%
T_{P}\approx 91.98K$, respectively. }
\label{figHP1}
\end{figure}
Provided that this scenario applies to cuprate superconductors, the
implications include: (i) the universal relations hold irrespective of the
dopant concentration and of the material; (ii) given the nonuniversal
critical amplitudes of the correlation lengths, $\xi _{i,0}^{\pm }$, and the
form of the universal scaling function $G_{3}^{\pm }(\mathcal{Z})$,
properties which can be derived from the free energy can be calculated close
to the zero field transition. These properties include the specific heat,
magnetic torque, diamagnetic susceptibility, melting line, etc. It should be
recognized that the universal relations, Eq. (\ref{HP4}), also imply
constraints on, e.g., isotope and pressure coefficients.

\begin{figure}
\centerline{\includegraphics[height=5.7cm]{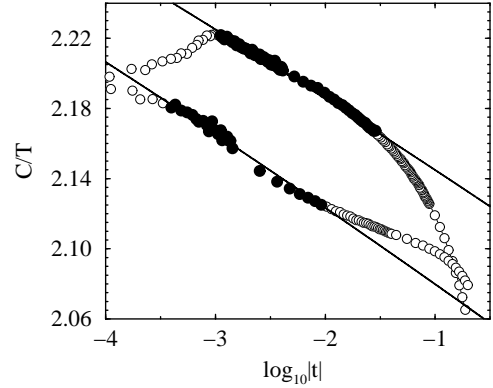}} \vskip -1.3cm
\caption{Specific heat coefficient $C/T$ $[mJ/(gK^{2})]$ versus $\log
_{10}|t|$ for $\mathrm{YBa_{2}Cu_{3}O_{7-\protect\delta }}$ (sample YBCO3,
\protect\cite{Charalambous:1999}) for $T_{c}=92.12K$. }
\label{figHP2}
\end{figure}
Although there is mounting evidence for $3D$-$XY$ universality in the
cuprates \cite
{SchneiderAriosa:1992,ToniHugo:1993,Schneider:1998_1,Hubbard:1996,Kamal:1994,Pas
ler:1998,Hofer:1998}
it appears impossible to prove that unambiguously. Indeed, due to
inhomogeneities, a solid always is homogeneous over a finite length $L$
only. In this case, the actual correlation length $\xi (t)\propto |t|^{-\nu }
$ cannot grow beyond $L$ as $t\rightarrow 0$, and the transition appears
rounded. Due to this finite size effect, the specific heat peak occurs at a
temperature $T_{P}$ shifted from the homogeneous system by an amount $%
L^{-1/\nu }$, and the magnitude of the peak located at temperature $T_{P}$
scales as $L^{\alpha /\nu }$. To quantify this point we show in Fig. \ref
{figHP1} the measured heat coefficient of
$\mathrm{YBa_{2}Cu_{3}O_{7-\delta }}$ \cite{Charalambous:1999} around the
peak.
The rounding and the shape of
the specific heat coefficient clearly exhibit the characteristic behavior of
a system in confined dimensions, i.e., rod or cube shaped inhomogeneities
\cite{Schultka:1995}. A finite size scaling analysis \cite{TSJMSPRL:1999}
reveals inhomogeneities with a characteristic length scale ranging from
$300$ to $400 \AA$, in the $\mathrm{YBa_{2}Cu_{3}O_{7-\delta}}$ samples YBCO3,
UBC2 and UBC1 of Ref. \cite{Charalambous:1999}. For this reason, deviations
from $3D$-$XY$ critical behavior around $T_{P}$ do not signal the failure of
$3D$-$XY$ universality, as previously claimed \cite{Charalambous:1999}, but
reflect a mere finite size effect at work. Indeed, from Fig. \ref{figHP2} it
is seen that the finite size effect makes it impossible to enter the
asymptotic critical regime. To set the scale we note that in the $\lambda $%
-transition of $^{4}$He the critical properties can be probed down to $%
|t|=10^{-9}$ \cite{Ahlers:1984,Metha:1999}. In Fig. \ref{figHP2} we marked
the intermediate regime where consistency with $3D$-$XY$ critical behavior,
i.e., with ${C}/{T}=\widetilde{A}^{\pm }10^{-\alpha \log _{10}|t|}+%
\widetilde{B}^{\pm }$ for $\alpha =-0.013$ and $\widetilde{A}^{+}/\widetilde{%
A}^{-}=1.07$, can be observed.
\begin{figure}[tbp]
\centerline{\includegraphics[height=5.7cm]{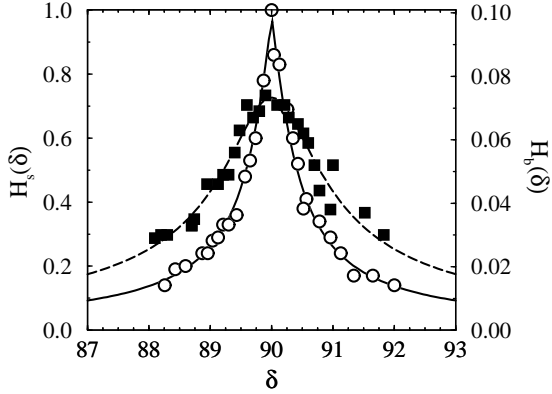}} \vskip -1cm
\caption{$H^{\ast }(\protect\delta )$ for $\mathrm{Bi_{2}Sr_{2}CaCu_{2}O_{8+%
\protect\delta }}$ with $T_{c}=84.1K$ for $T=79.5K$ ($\circ $) and $T=82.8K$
($\blacksquare $). Experimental data are taken from \protect\cite{Roemer},
the solid lines are derived from the respective $2D$ ($H_{s}$, `slab', ---)
and $3D$ ($H_{b}$, `bulk' with $\protect\gamma =77$, - - -) scaling forms. }
\label{figHP6}
\end{figure}
\begin{figure}
\centerline{\includegraphics[height=5.2cm]{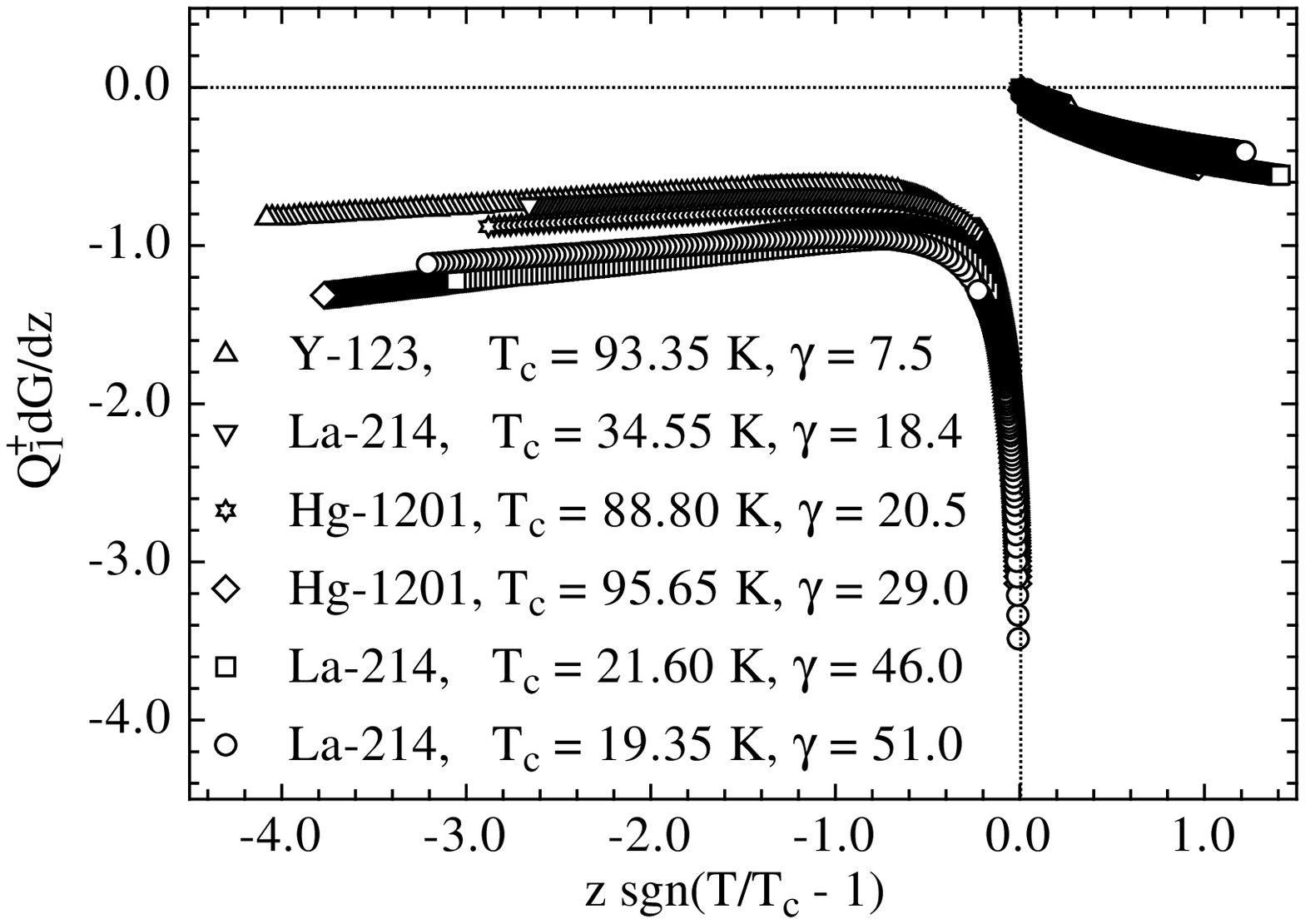}} \vskip -1.1cm
\caption{\mbox{Scaling function $dG_{3}^{\pm }(\mathcal{Z})/d\mathcal{Z}$
derived}
from the angular dependence of the magnetic
torque for $\mathrm{YBa_{2}Cu_{3}O_{6.93}}$,
$\mathrm{La_{1.854}Sr_{0.146}CuO_{4}}$,
\mbox{$\mathrm{HgBa_{2}CuO_{4.108}}$, $\mathrm{HgBa_{2}CuO_{4.096}}$,}
$\mathrm{La_{1.914}Sr_{0.086}CuO_{4}}$ and $\mathrm{%
La_{1.920}Sr_{0.080}CuO_{4}}$.}
\label{figHP3}
\end{figure}
The upper branch corresponds to $T<T_{c}$ and the lower one to $T>T_{c}$.
The open circles closer to $T_{c}$ correspond to the finite size affected
region, while further away the temperature dependence of the background,
usually attributed to phonons, becomes significant. Hence, due to the finite
size effect and the temperature dependence of the background the
intermediate regime is bounded by the temperature region where the data
depicted in Fig. \ref{figHP2} fall nearly on straight lines. To provide
quantitative evidence for $3D$-$XY$ universality in this regime, we invoke
the universal relation (\ref{HP4}) and calculate $T_{c}$ from the critical
amplitudes of specific heat and penetration depth. Using $A^{+}=8.4\cdot
10^{20}cm^{-3}$, derived from the data shown in Fig. \ref{figHP2} for sample
YBCO3 with $T_{c}=92.12K$, $\lambda _{a,0}=1153\AA $, $\lambda _{b,0}=968\AA
$ and $\lambda _{c,0}=8705\AA $, derived from magnetic torque measurements
on a sample with $T_{c}=91.7K$ \cite{Schneider:1998_1}, as well as the
universal numbers $A^{+}/A^{-}=1.07$ and $\mathcal{R}^{-}\approx 0.59$, we
obtain $T_{c}=88.2K$. Hence, the universal $3D$-$XY$ relation (\ref{HP4}) is
remarkably well satisfied.

Another difficulty results from the pronounced anisotropy of the cuprates: a
convenient measure is the effective mass parameter $\gamma =\sqrt{%
M_{\Vert}/M_{\bot }}$, which depends on the dopant concentration. Even
though the strength of thermal fluctuations grows with increasing $\gamma $,
they are slightly away from $T_{c}$ essentially two-dimensional.
Accordingly, the intermediate $3D$-$XY$ critical regime shrinks, and the
corrections to scaling are expected to become significant. An experimental
demonstration of the temperature driven dimensional crossover is shown in
Fig. \ref{figHP6} in terms of the angular dependence of the onset field $H^*$%
, where a measurable resistance is observed \cite{Roemer}. $H^*(\delta)$
follows from $\mathcal{Z}(H^*)=\mathcal{Z}^*$. For superconducting sheets of
thickness $d_s$, corresponding to $2D$, the argument of the scaling function
$G_2^\pm(\mathcal{Z})$ is given by \cite{SchneiderTanner,OOI} {\small
\begin{equation}
\mathcal{Z}=\left(\frac{H}{\Phi_0}(\xi_\parallel^\pm)^2\vert\cos(\delta)%
\vert +\frac{H^2}{\Phi_0^2}(\xi_\parallel^\pm)^2d_s^2\sin^2(\delta)%
\right)^{1/2}.  \label{HP10}
\end{equation}
}
\begin{figure}
\vskip 0.2cm
\centerline{\includegraphics[height=5.2cm]{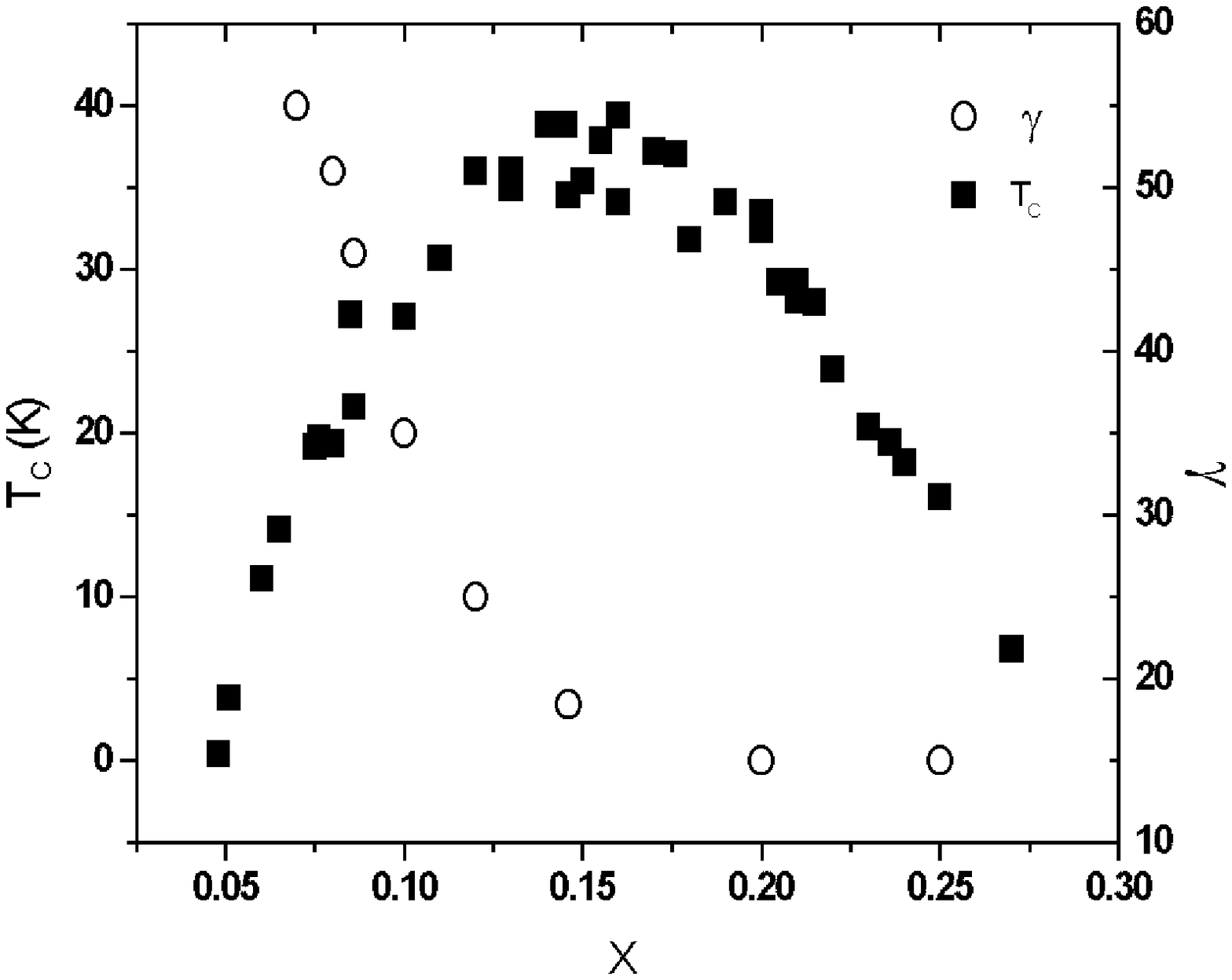}} \vskip -0.8cm
\caption{$T_{c}$ and $\protect\gamma=\protect\sqrt{M_{\bot }/M_{\Vert }}$
versus dopant concentration $x$ for $\mathrm{La_{2-x}Sr_{x}CuO_{4}}$. }
\label{figHP4}
\end{figure}
Noting that Eqs. (\ref{HP7}) and (\ref{HP10}) lead to distinct bell-shaped
($3D$) and cusp-like ($2D$) behavior around $\delta=90^o$, respectively,
these measurements clearly illustrate the temperature driven dimensional
crossover. Indead, as seen in Fig. \ref{figHP6}, at $T=79.5K$ $H^*(\delta)$
mirrors a $2D$ film behavior, while closer to $T_c$ at $T=82.8K$ $3D$ bulk
behavior appears. To illustrate the difficulties associated with this
crossover we show in Fig. \ref{figHP3} estimates for the derivative of the
universal scaling function $G_{3}^{\pm }(\mathcal{Z})$ derived from magnetic
torque measurements \cite{Hofer:1998}. Even though the qualitative behavior
is the same for all samples, the deviations increase systematically with
increasing $\gamma$. This systematics cannot be attributed to the
experimental uncertainties of about $40\%$. It is more likely that it
reflects the $3D$ to $2D$ crossover and the associated reduction of the $3D$-%
$XY$ fluctuation dominated regime, requiring corrections to scaling. Indeed,
in the derivation of the scaling function from the experimental data, both,
corrections to scaling and finite size effects have not been considered.

In materials, such as $\mathrm{La_{2-x}Sr_{x}CuO_{4}}$, where the underdoped
regime is experimentally accessible, a strict $3D$- to $2D$-$XY$ crossover
occurs. By approaching the underdoped limit $x=x_{u}\approx 0.05$, $\gamma =%
\sqrt{M_{\bot }/M_{\Vert }}$ becomes very large (see Fig. \ref{figHP4}) and $%
T_{c}$ vanishes. Here the materials correspond to a stack of independent
sheets of thickness $d_{s}$ As there is a phase transition line with an
endpoint $T_{c}(x=x_{u})=0$, one expects a doping driven $2D$-$XY$ insulator
to superconductor transition at $T=0$.
\begin{figure}
\centerline{\setlength{\unitlength}{1cm}
\begin{picture}(8.5,5.5)
\put(0.2,-1.){\includegraphics[height=6.0cm]{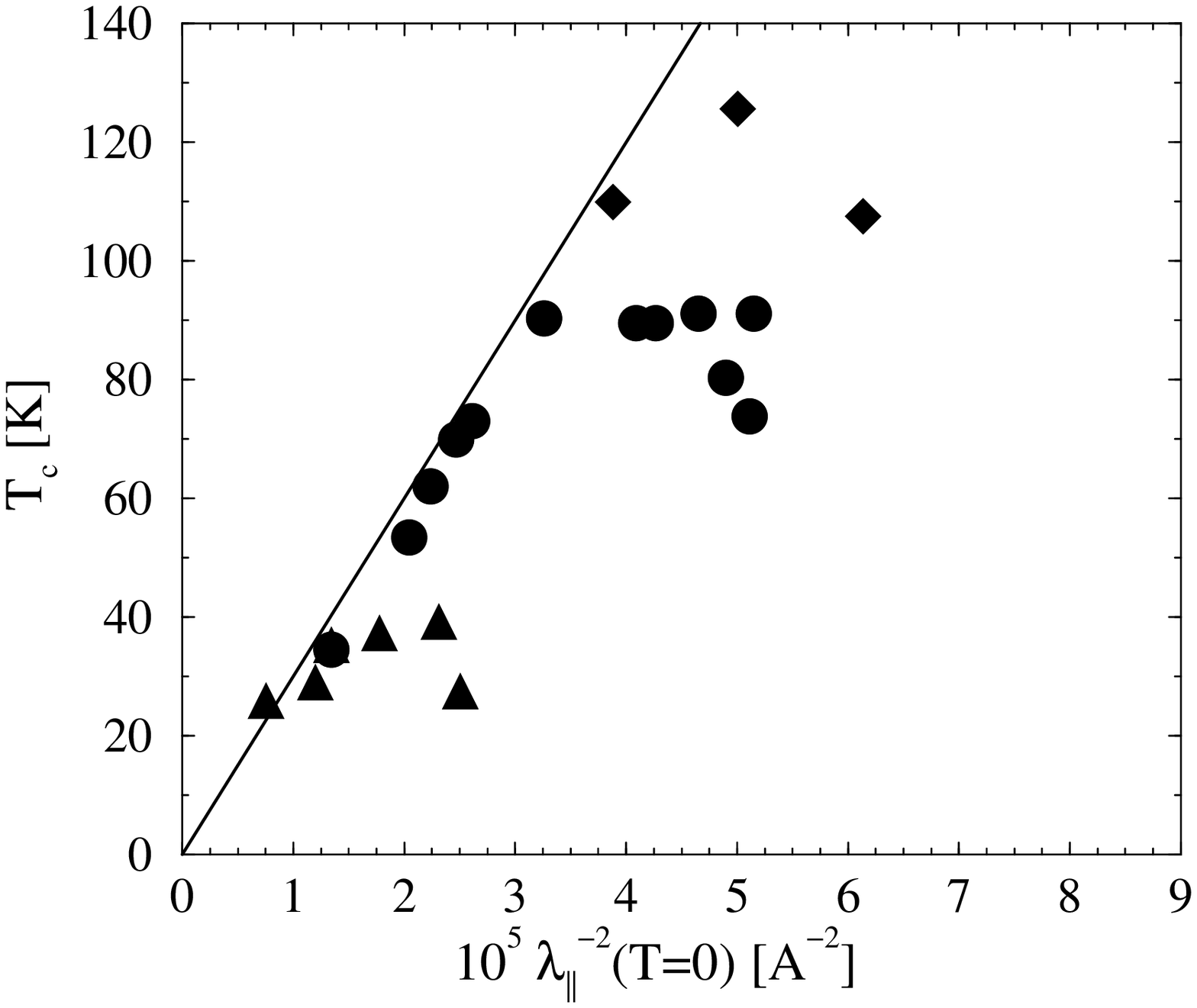}}
\put(3.3,0.05){\includegraphics[height=3.1cm]{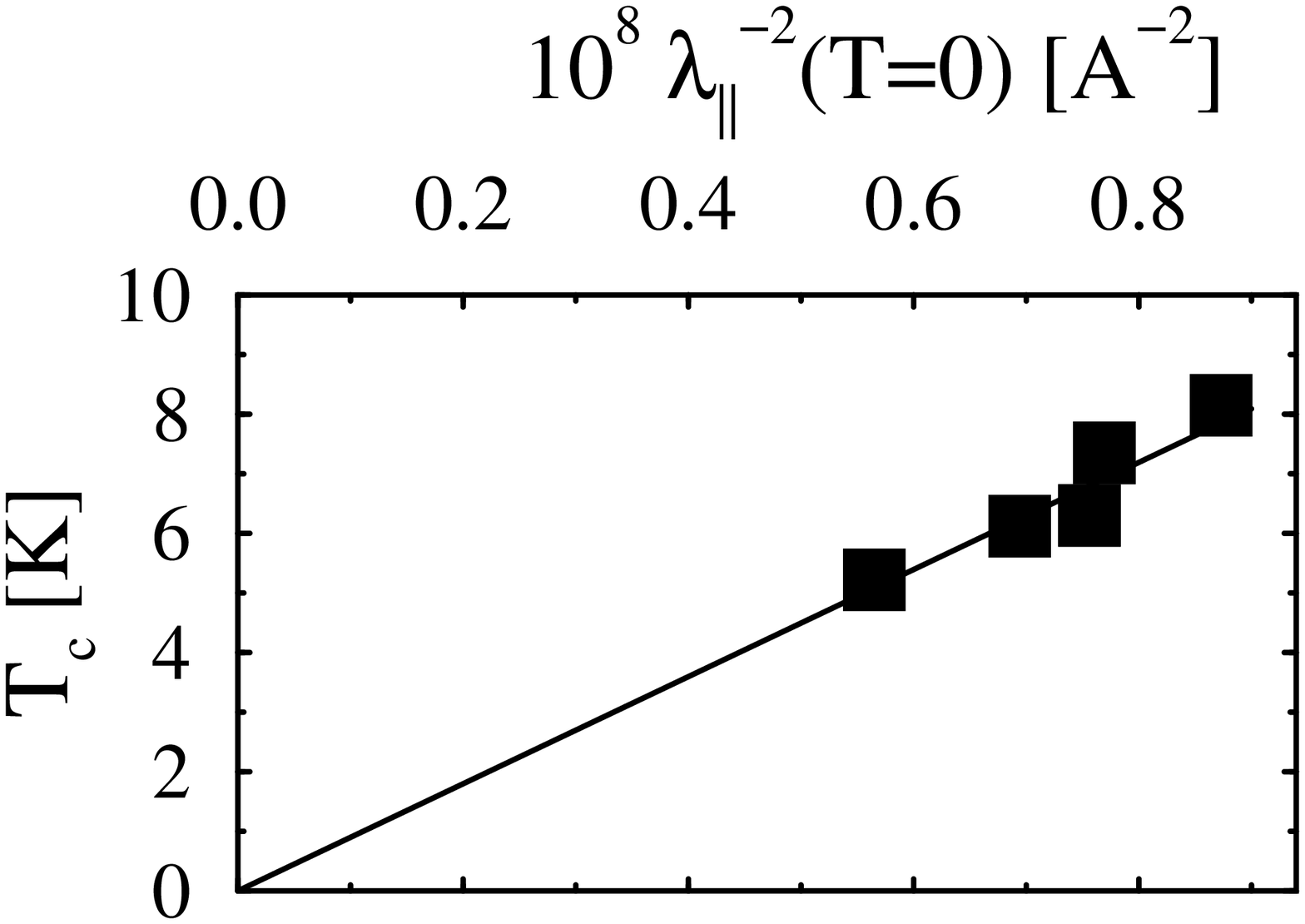}}
\end{picture}} \vglue -0.1cm
\caption{$T_{c}$ versus $\protect\lambda_{\parallel }^{-2}(T\rightarrow 0)$
as obtained from $\protect\mu$SR measurements for $\mathrm{%
La_{2-x}Sr_{x}CuO_{4}}$ ($\blacktriangle $), $\mathrm{YBa_{2}Cu_{3}O_{7-x}}$%
, $\mathrm{Bi_{2}Sr_{2}Ca_{1-x}Y_{x}Cu_{2}O_{8+\protect\delta }}$ ($\bullet $%
), and $\mathrm{Tl_{2}Ba_{2}Ca_{2}Cu_{3}O_{10}}$ ($\blacklozenge $). Data
taken from \protect\cite{Uemura:1987}. The straight line marks $T_{c}=3\cdot
10^{8}\protect\lambda _{ \parallel }^{-2}$ with $T_{c}$ in $[K]$ and $%
\protect\lambda_{\parallel }$ in $[\AA]$. Inset: $T_{c}$ vs. $\protect%
\lambda _{\parallel }^{-2}(T=0)$ of $\mathrm{Bi_{2+x}Sr_{2-x}CuO_{6+\protect%
\delta }}$ ($\blacksquare $); the straight line is a guide to the eye. Data
taken from \protect\cite{Janod:1996}. }
\label{figHP5}
\end{figure}
For such a transition the scaling theory of quantum critical phenomena
predicts \cite{Cha:1991,Kim:1991,Schneider:1996} {\small
\begin{eqnarray}
\lim_{\delta \rightarrow 0}\frac{1}{d_{s}^{2}}{(}T{_{c}(\delta ))^{2}\lambda
_{x}^{2}(\delta ,}T{=0)\lambda _{y}^{2}(\delta ,}T{=0)}=  \notag \\
=\frac{1}{Q_{2,0}^{2}} \left( {\frac{\Phi _{0}^{2}}{16\pi ^{3}k_{B}}}\right)
^{2}  \label{HP8}
\end{eqnarray}
} to be universal. $\delta =(x-x_{u})/x_{u}$ is the control parameter, $%
Q_{2,0}$ is an universal number and {\small
\begin{equation}
T_{c}\propto\delta ^{z\overline{\nu }},\quad {\lambda _{i}^{2}(\delta ,}T{=0)%
}\propto \delta ^{-\overline{\nu }}.  \label{HP9}
\end{equation}
} $z$ is the dynamic critical exponent and $\overline{\nu }$ the exponent of
the correlation length. In Fig. \ref{figHP5} we depict experimental data in
terms of $T_{c}$ versus $1/\lambda _{\Vert }^{2}(T=0)$. As $T_c$ approaches
the underdoped limit, the data appear to merge on the solid line. In this
context it should be emphasized that the data, with the exception of $%
\mathrm{Bi_{2+x}Sr_{2-x}CuO_{6+\delta}}$, are rather far away from the
asymptotic regime where Eq. (\ref{HP8}) is expected to apply. Moreover, $%
d_{s}$ is known to adopt material dependent values \cite{TSJMS:1999}.
Nevertheless, the data collected in Figs. \ref{figHP4} and \ref{figHP5}
clearly point to a quantum phase transition in $D=2$ at $x=x_{u}$ where $%
T_{c}$ vanishes and $\lambda _{\Vert }^{2}(T=0)$ tends to infinity. Hence,
there is strong evidence for a $2D$-$XY$ quantum phase transition, where Eq.
(\ref{HP8}) applies and $dT_{c}/d(1/\lambda_{\Vert}^{2}(T=0))$ is not
universal, as suggested by Uemura et al \cite{Uemura:1987}, but depends on $%
d_{s}$.

To summarize, there is mounting evidence for intermediate $D$-$XY$ critical
behavior, a $3D$- to $2D$- crossover as the underdoped limit is approached
and for the occurrence of a quantum superconductor to insulator transition
at the underdoped limit in $D=2$. Emerging implications and constraints for
microscopic models include: (i) in the experimentally accessible temperature
regime and close to optimum doping, there is remarkable consistency with $3D$%
-$XY$ universality; (ii) close to criticality the symmetry of the order
parameter is either d-wave or s-wave; (iii) the decrease of $T_{c}$ in the
underdoped regime mirrors the dimensional crossover, enhancing thermal
fluctuations and the competition with quantum fluctuations which suppress
superconductivity at the underdoped limit; (iv) the enhanced thermal and
quantum fluctuations reduce the single particle density of states at the
chemical potential. This reduction leads to a pseudogap above $T_{c}$; (v)
these fluctuations imply the existence of phase uncorrelated pairs above $%
T_{c}$ and invalidate mean-field treatments, including the Fermi liquid
approach in the normal state; etc.

For a more elaborate review of the $D$-$XY$ behavior in cuprate
superconductors we refer to Ref. \cite{BOOK}.

The authors are grateful to J. Hofer for very useful comments and
suggestions on the subject matter.

\end{document}